\documentstyle[12pt,epsf,psfig,graphicx]{article}

\begin{document}

\title{{\bf Intermittency at critical transitions and aging dynamics at edge of
chaos}}
\author{A. Robledo \\
Instituto de F\'{i}sica,\\
Universidad Nacional Aut\'{o}noma de M\'{e}xico,\\
Apartado Postal 20-364, M\'{e}xico 01000 D.F., Mexico.}
\date{.}
\maketitle

\begin{abstract}
We recall that, at both the intermittency transitions and at the Feigenbaum
attractor in unimodal maps of non-linearity of order $\zeta >1$, the
dynamics rigorously obeys the Tsallis statistics. We account for the $q$%
-indices and the generalized Lyapunov coefficients $\lambda _{q}$ that
characterize the universality classes of the pitchfork and tangent
bifurcations. We identify the Mori singularities in the Lyapunov spectrum at
the edge of chaos with the appearance of a special value for the entropic
index $q$. The physical area of the Tsallis statistics is further probed by
considering the dynamics near criticality and glass formation in thermal
systems. In both cases a close connection is made with states in unimodal
maps with vanishing Lyapunov coefficients.


\end{abstract}

\section{Introduction}

Ever since the proposition, in 1988, by Tsallis \cite{tsallis0}, \cite
{tsallis1} of a nonextensive generalization of the canonical formalism of
statistical mechanics a spirited discussion \cite{cho1}, \cite{latora1} has
grown on the foundations of this branch of physics. But, only until recently 
\cite{robledo0}-\cite{robledo3} has there appeared firm evidence about the
relevance of the generalized formalism for specific model system situations.
These studies present rigorous results on the dynamics associated to
critical attractors in prototypical nonlinear one-dimensional maps, such as
those at the pitchfork and tangent bifurcations and at the accumulation
point of the former, the so-called onset of chaos \cite{schuster1}. As these
are very familiar and well understood features of these maps it is of
interest to see how previous knowledge fits in with the new perspective.
Also, clear-cut calculations may help clarify the physical reasons (believed
in these examples to be a breakdown in the chain of increasing randomness
from non-ergodicity to completely developed chaoticity) for the departure
from the Boltzmann-Gibbs (BG) statistics and the competence of the
non-extensive generalization. Here we review briefly specific results
associated to the aforementioned states in unimodal maps which are all
characterized by vanishing Lyapunov coefficients. We also recall how the
dynamics at the tangent bifurcation appears to be related to that at thermal
critical states. In addition, time evolution at the noise-perturbed onset of
chaos is seen to be closely analogous to the glassy dynamics observed in
supercooled molecular liquids.

\section{Dynamics at the pitchfork and tangent bifurcations}

As pointed out in Refs. \cite{robledo0}, \cite{baldovin1}, \cite{robledo1}
the long-known \cite{schuster1} exact geometric or {\it static} solution of
the Feigenbaum renormalization group (RG) equations for the tangent
bifurcation in unimodal maps of nonlinearity $\zeta >1$ also describes the 
{\it dynamics} of iterates at such state. A straightforward extension of
this approach applies also to the pitchfork bifurcations. We recall that the
period-doubling and intermittency transitions are based on the pitchfork and
the tangent bifurcations, respectively, and that at these critical states
the ordinary Lyapunov coefficient $\lambda _{1}$ vanishes. The sensitivity
to initial conditions $\xi _{t}$ was determined analytically and its
relation with the rate of entropy production examined \cite{robledo1}. The
fixed-point expressions have the specific form that corresponds to the
temporal evolution suggested by the nonextensive formalism. These studies
contain the derivation of their $q$-generalized Lyapunov coefficients $%
\lambda _{q}$ and the description of the different possible types of
sensitivity $\xi _{t}$.

By considering as starting point the $\zeta $-logistic map $f_{\mu
}(x)=1-\mu \left| x\right| ^{\zeta }$,$\;\zeta >1$, $-1\leq x\leq 1$, it is
found that for both infinite sets of pitchfork and tangent bifurcations $\xi
_{t}$, defined as $\xi _{t}\equiv \lim_{\Delta x_{0}\to 0}(\Delta
x_{t}/\Delta x_{0})$ (where $\Delta x_{0}$ is the length of the initial
interval and $\Delta x_{t}$ its length at time $t$), has the form suggested
by Tsallis, 
\begin{equation}
\xi _{t}(x_{0})=\exp _{q}[\lambda _{q}(x_{0})\ t]\equiv [1-(q-1)\lambda
_{q}(x_{0})\ t]^{-\frac{1}{q-1}},  \label{sensitivity_00}
\end{equation}
that yields the customary exponential $\xi _{t}$ with Lyapunov coefficient $%
\lambda _{1}(x_{0})$ when $q\rightarrow 1$. In Eq. (\ref{sensitivity_00}) $q$
is the entropic index and $\lambda _{q}$ is the $q$-generalized Lyapunov
coefficient; $\exp _{q}(x)\equiv [1-(q-1)x]^{-1/(q-1)}$ is the $q$%
-exponential function. The pitchfork and the left-hand side of the tangent
bifurcations display weak insensitivity to initial condition, while the
right-hand side of the tangent bifurcations presents a `super-strong'
(faster than exponential) sensitivity to initial conditions \cite{baldovin1}.

For the transition to periodicity of order $n$ the composition $f_{\mu
}^{(n)}$ is first considered. In the neighborhood of one of the $n$ points
tangent to the line with unit slope one obtains $f^{(n)}(x)=x+u\left|
x\right| ^{z}+o(\left| x\right| ^{z})$, where $u>0$ is the expansion
coefficient. The general result obtained is $q=2-z^{-1}$ and $\lambda
_{q}(x_{0})=zux_{0}^{z-1}$\cite{baldovin1}, \cite{robledo1}. At the tangent
bifurcations one has $f_{\mu }^{(n)}(x)=x+ux^{2}+o(x^{2})$,$\ u>0$, and from 
$z=2$ one gets $q=3/2$. For the pitchfork bifurcations one has instead $%
f_{\mu }^{(n)}(x)=x+ux^{3}+o(x^{3})$, because $d^{2}f_{\mu
}^{(2^{k})}/dx^{2}=0$ at these transitions, and $u<0$ is now the coefficient
associated to $d^{3}f_{\mu }^{(2^{k})}/dx^{3}<0$. In this case we have $z=3$
in $q=2-z^{-1}$ and one obtains $q=5/3$. Notably, these specific results for
the index $q$ are valid for all $\zeta >1$ and therefore define the
existence of only two universality classes for unimodal maps, one for the
tangent and the other one for the pitchfork bifurcations \cite{baldovin1}.
See Figs. 2 and 3 in Ref. \cite{baldovin1}.

Notice that our treatment of the tangent bifurcation differs from other
studies of intermittency transitions \cite{gaspard1} in that there is no
feed back mechanism of iterates into the origin of $f^{(n)}(x)$ or of its
associated fixed-point map. Here impeded or incomplete mixing in phase space
(a small interval neighborhood around $x=0$) arises from the special
'tangency' shape of the map at the pitchfork and tangent transitions that
produces monotonic trajectories. This has the effect of confining or
expelling trajectories causing anomalous phase-space sampling, in contrast
to the thorough coverage in generic states with $\lambda _{1}>0$. By
construction the dynamics at the intermittency transitions, describe a
purely nonextensive regime.

\section{Link between intermittent dynamics and dynamics at thermal
criticality}

An unanticipated relationship has been shown by Contoyiannis {\it et al} 
\cite{athens1}-\cite{athens4} to exist between the intermittent dynamics of
nonlinear maps at a tangent bifurcation and the equilibrium dynamics of
fluctuations at an ordinary thermal critical point. With the aim of
obtaining the properties of clusters or domains of the order parameter at
criticality, a Landau-Ginzburg-Wilson (LGW) approach was carried out such
that the dominant contributions to the partition function arise from a
singularity (similar to an instanton) located in the space outside the
cluster \cite{athens1}, \cite{athens2}. Then \cite{athens3}, \cite{athens4},
a nonlinear map for the average order parameter was constructed whose
dynamics reproduce the averages of the cluster critical properties. This map
has as a main feature the tangent bifurcation and as a result time evolution
is intermittent.

The starting point is the partition function of the $d$-dimensional system
at criticality, 
\begin{equation}
Z=\int D[\phi ]\exp (-\Gamma _{c}[\phi ]),  \label{partition1}
\end{equation}
where 
\begin{equation}
\Gamma _{c}[\phi ]=g_{1}\int_{\Omega }dV\left[ \frac{1}{2}(\nabla \phi
)^{2}+g_{2}\left| \phi \right| ^{\delta +1}\right]  \label{landau1}
\end{equation}
is the critical LGW free energy of a system of $d$-dimensional volume $%
\Omega $, $\phi $ is the order parameter (e.g. magnetization per unit
volume) and $\delta $ is the critical isotherm exponent. By considering the
space-averaged magnetization $\Phi =\int_{V}\phi (x)dV$, the statistical
weight 
\begin{equation}
\rho (\Phi )=\exp (-\Gamma _{c}[\Phi ])/Z,  \label{invariant1}
\end{equation}
where $\Gamma _{c}[\Phi ]\sim g_{1}g_{2}\Phi ^{\delta +1}$ and $Z=\int d\Phi
\exp (-\Gamma _{c}[\Phi ])$, was shown to be the invariant density of a
statistically equivalent one-dimensional map. The functional form of this
map was obtained as the solution of an inverse Frobenius-Perron problem \cite
{athens3}. For small values of $\Phi $ the map has the form 
\begin{equation}
\Phi _{n+1}=\Phi _{n}+u\Phi _{n}^{\delta +1}+\epsilon ,  \label{tangent1}
\end{equation}
where the amplitude $u$ depends on $g_{1}$, $g_{2}$and $\delta $, and the
shift parameter $\epsilon \sim R^{-d}$. Eq. (\ref{tangent1}) can be
recognized as that describing the intermittency route to chaos in the
vicinity of a tangent bifurcation \cite{schuster1}. The complete form of the
map displays a superexponentially decreasing region that takes back the
iterate close to the origin in one step. Thus the parameters of the thermal
system determine the dynamics of the map. Averages made of order-parameter
critical configurations are equivalent to iteration time averages along the
trajectories of the map close to the tangent bifurcation. The mean number of
iterations in the laminar region was seen to be related to the mean
magnetization within a critical cluster of radius $R$. There is a
corresponding power law dependence of the duration of the laminar region on
the shift parameter $\epsilon $ of the map \cite{athens3}. For $\epsilon >0$
the (small) Lyapunov coefficient is simply related to the critical exponent $%
\delta $ \cite{athens2}.

As the size of subsystems or domains of the critical system is allowed to
become infinitely large the Lyapunov coefficient vanishes and the duration
of the laminar episodes of intermittency diverge. Without the feedback to
the origin feature in the map we recover the conditions described in the
previous section. See Ref. \cite{robledo2} for more details.

\section{ Dynamics inside the Feigenbaum attractor}

The dynamics at the chaos threshold, also referred to as the Feigenbaum
attractor, of the $\zeta $-logistic map at $\mu _{c}$ has been analyzed
recently \cite{baldovin2}, \cite{baldovin3}. By taking as initial condition $%
x_{0}=0$ we found that the resulting orbit consists of trajectories made of
intertwined power laws that asymptotically reproduce the entire
period-doubling cascade that occurs for $\mu <\mu _{c}$. This orbit captures
the properties of the so-called 'superstable' orbits at $\overline{\mu }%
_{n}< $ $\mu _{c}$, $n=1,2,...$ \cite{schuster1}. Here again the Lyapunov
coefficient $\lambda _{1}$ vanishes and in its place there appears a
spectrum of $q$-Lyapunov coefficients $\lambda _{q}^{(k)}$. This spectrum
was originally studied in Refs. \cite{politi1}, \cite{mori1} and our
interest has been to examine the relationship of its properties with the
Tsallis statistics. We found that the sensitivity to initial conditions has
precisely the form of a $q$-exponential, of which we determine the $q$-index
and the associated $\lambda _{q}^{(k)}$. The appearance of a specific value
for the $q$ index (and actually also that for its conjugate value $Q=2-q$)
turns out to be due to the occurrence of Mori's '$q$ transitions' \cite
{mori1} between 'local attractor structures' at $\mu _{c}$. Furthermore, we
have also shown that the dynamical and entropic properties at $\mu _{c}$ are
naturally linked through the nonextensive expressions for the sensitivity to
initial conditions $\xi _{t}$ and for the entropy $S_{q}$ in the rate of
entropy production $K_{q}^{(k)}$. We have corroborated analytically \cite
{baldovin2}, the equality $\lambda _{q}^{(k)}=$ $K_{q}^{(k)}$ given by the
nonextensive statistical mechanics. Our results support the validity of the $%
q$-generalized Pesin identity at critical points of unimodal maps.

\begin{figure}[htbp]
\centering
\includegraphics[width=6cm ,angle=-90]{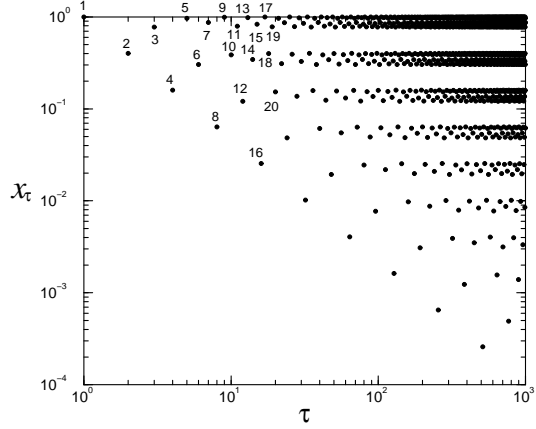}
\caption{Absolute values of positions in logarithmic scales of the first $%
1000$ iterations $\tau $ for a trajectory of the logistic map at the onset
of chaos $\mu _{c}(0)$ with initial condition $x_{in}=0$. The numbers
correspond to iteration times. The power-law decay of the time subsequences
described in the text can be clearly appreciated. }
\label{figura1}
\end{figure}

Thus, the absolute values for the positions $x_{\tau }$ of the trajectory
with $x_{t=0}=0$ at time-shifted $\tau =t+1$ have a structure consisting of
subsequences with a common power-law decay of the form $\tau ^{-1/1-q}$ with 
$q=1-\ln 2/(\zeta -1)\ln \alpha (\zeta )$, where $\alpha (\zeta )$ is the
Feigenbaum universal constant that measures the period-doubling
amplification of iterate positions \cite{baldovin1}. That is, the Feigenbaum
attractor can be decomposed into position subsequences generated by the time
subsequences $\tau =(2k+1)2^{n}$, each obtained by proceeding through $%
n=0,1,2,...$ for a fixed value of $k=0,1,2,...$. See Fig. \ref{figura1}. The 
$k=0$  subsequence can be written as $x_{t}=\exp _{2-q}(-\lambda _{q}^{(0)}t)
$ with $\lambda _{q}^{(0)}=(\zeta -1)\ln \alpha (\zeta )/\ln 2$. These
properties follow from the use of $x_{0}=0$ in the scaling relation \cite
{baldovin1} 
\begin{equation}
x_{\tau }\equiv \left| g^{^{(\tau )}}(x_{0})\right| =\tau ^{-1/1-q}\left|
g(\tau ^{1/1-q}x_{0})\right| .  \label{trajectory1}
\end{equation}

The sensitivity associated to trajectories with other starting points $%
x_{0}\neq 0$ within the attractor can be determined similarly with the use
of the time subsequences $\tau =(2k+1)2^{n}$. One obtains $\lambda
_{q}^{(k)}=(\zeta -1)\ln \alpha (\zeta )/(2k+1)\ln 2>0$, $k=0,1,2,...$, the
positive branch of the Lyapunov spectrum, when the trajectories start at the
most crowded ($x_{\tau =0}=1$) and finish at the most sparse ($x_{\tau
=2^{n}}=0$) region of the attractor. By inverting the situation we obtain $%
\lambda _{2-q}^{(k)}=-2(\zeta -1)\ln \alpha /(2k+1)\ln 2<0$, $k=0,1,2,...$,
the negative branch of $\lambda _{q}^{(k)}$, i.e. starting at the most
sparse ($x_{\tau =0}=0$) and finishing at the most crowded ($x_{\tau
=2^{n}+1}=0$) region of the attractor. Notice that $Q=2-q$ as $\exp
_{Q}(y)=1/\exp _{q}(-y)$. For the case $\zeta =2$ see Refs. \cite{baldovin2}
and \cite{baldovin3}, for general $\zeta >1$ see Refs. \cite{mayoral1} and 
\cite{mayoral2} where also a different and more direct derivation is used.
So, when considering these two dominant families of orbits all the $q$%
-Lyapunov coefficients appear associated to only two specific values of the
Tsallis index, $q$ and $2-q$.

\begin{figure}[htbp]
\centering
\includegraphics[width=10cm ,angle=0]{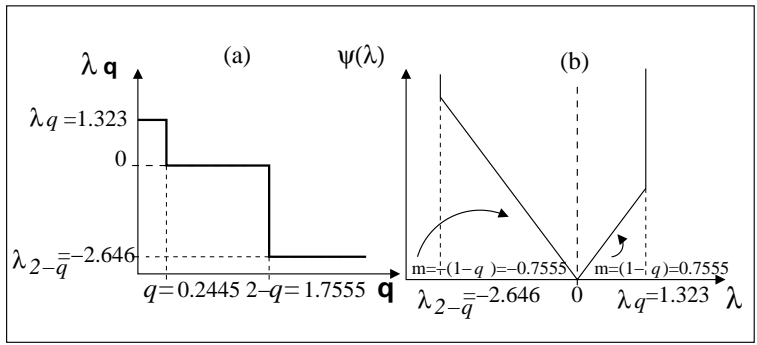}
\caption{a) The Lyapunov coefficient function $\lambda({\sf q})$ at the
chaos threshold at $\mu _{c}$ and b) the spectrum $\psi (\lambda )$. See
text for description}
\label{figura2}
\end{figure}

As a function of the running variable $-\infty <{\sf q}<\infty $ the $%
\lambda _{q}^{(k)}$ coefficients become a function $\lambda ({\sf q})$ with
two steps located at ${\sf q}=q=1\mp \ln 2/(\zeta -1)\ln \alpha (\zeta )$.
In this manner contact can be established with the formalism developed by
Mori and coworkers and the $q$ phase transition obtained in Ref. \cite{mori1}%
. The step function for $\lambda ({\sf q})$ can be integrated to obtain the
spectrum $\phi ({\sf q})$ ($\lambda ({\sf q})\equiv d\phi /d\lambda ({\sf q}%
) $) and its Legendre transform $\psi (\lambda )$ ($\equiv \phi -(1-{\sf q}%
)\lambda $), the dynamic counterparts of the Renyi dimensions $D_{{\sf q}}$
and the spectrum $f(\alpha )$ that characterize the geometry of the
attractor. The constant slopes in the spectrum $\psi (\lambda )$ represent
the Mori's $q$ transitions for this attractor and the value $1-q$ coincides
with that of the slope previously detected \cite{politi1}, \cite{mori1}. 
See Fig. \ref{figura2}. Details appear in Ref. \cite{mayoral2}.

Ensembles of trajectories with starting points close to $x_{\tau =0}=1$
expand in such a way that a uniform distribution of initial conditions
remains uniform for all later times $t\leq T$ where $T$ marks the crossover
to an asymptotic regime. As a consequence of this we established \cite
{baldovin3} the identity of the rate of entropy production $K_{q}^{(k)}$
with $\lambda _{q}^{(k)}$. The $q$-generalized rate of entropy production $%
K_{q}$ is defined via $K_{q}t=S_{q}(t)-S_{q}(0)$, $t$ large, where 
\begin{equation}
S_{q}\equiv \sum_{i}p_{i}\ln _{q}\left( \frac{1}{p_{i}}\right) =\frac{%
1-\sum_{i}^{W}p_{i}^{q}}{q-1}  \label{tsallis1}
\end{equation}
is the Tsallis entropy, and where $\ln _{q}y\equiv (y^{1-q}-1)/(1-q)$ is the
inverse of $\exp _{q}(y)$. See Figs. 2 and 3 in Ref. \cite{baldovin3}.

Consider now the logistic map $\zeta $ $=2$ in the presence of additive
noise 
\begin{equation}
x_{t+1}=f_{\mu }(x_{t})=1-\mu x_{t}^{2}+\chi _{t}\sigma ,\;-1\leq x_{t}\leq
1,0\leq \mu \leq 2,  \label{logistic1}
\end{equation}
where $\chi _{t}$ is Gaussian-distributed with average $\left\langle \chi
_{t}\chi _{t^{\prime }}\right\rangle =\delta _{t.t^{\prime }}$, and $\sigma $
measures the noise intensity. We recall briefly the known properties of this
problem \cite{schuster1}, \cite{crutchfield1}. Except for a set of zero
measure, all the trajectories with $\mu _{c}(\sigma =0)$ and initial
condition $-1\leq x_{0}\leq 1$ fall into the attractor with fractal
dimension $d_{f}=0.5338...$. These trajectories represent nonergodic states,
since as $t\rightarrow \infty $ only a Cantor set of positions is accessible
out of the total phase space $-1\leq x\leq 1$. For $\sigma >0$ the noise
fluctuations wipe the sharp features of the periodic attractors as these
widen into bands similar to those in the chaotic attractors, nevertheless
there remains a well-defined transition to chaos at $\mu _{c}(\sigma )$
where the Lyapunov exponent $\lambda _{1}$ changes sign. The period doubling
of bands ends at a finite value $2^{N(\sigma )}$ as the edge of chaos
transition is approached and then decreases at the other side of the
transition. This effect displays scaling features and is referred to as the
bifurcation gap \cite{schuster1}, \cite{crutchfield1}. When $\sigma >0$ the
trajectories visit sequentially a set of $2^{n}$ disjoint bands or segments
leading to a cycle, but the behavior inside each band is completely chaotic.
These trajectories represent ergodic states as the accessible positions have
a fractal dimension equal to the dimension of phase space. Thus the removal
of the noise $\sigma \rightarrow 0$ leads to an ergodic to nonergodic
transition in the map.

In the presence of noise ($\sigma $ small) one obtains instead of Eq. (\ref
{trajectory1}) \cite{robledo3} 
\begin{equation}
x_{\tau }=\tau ^{-1/1-q}\left| g(\tau ^{1/1-q}x)+\chi \sigma \tau
^{1/1-r}G_{\Lambda }(\tau ^{1/1-q}x)\right| ,  \label{trajectory3}
\end{equation}
where $G_{\Lambda }(x)$ is the first order perturbation eigenfunction, and
where $r=1-\ln 2/\ln \kappa \simeq 0.6332$. Use of $x_{0}=0$ yields $x_{\tau
}=\tau ^{-1/1-q}\left| 1+\chi \sigma \tau ^{1/1-r}\right| $ or $x_{t}=\exp
_{2-q}(-\lambda _{q}t)\left[ 1+\chi \sigma \exp _{r}(\lambda _{r}t)\right] $
where $t=\tau -1$ and $\lambda _{r}=\ln \kappa /\ln 2$. At each noise level $%
\sigma $ there is a 'crossover' or 'relaxation' time $t_{x}=\tau _{x}-1$
when the fluctuations start suppressing the fine structure of the orbits
with $x_{0}=0$. This time is given by $\tau _{x}=\sigma ^{r-1}$, the time
when the fluctuation term in the perturbation expression for $x_{\tau }$
becomes unbounded by $\sigma $, i.e. $x_{\tau _{x}}=\tau _{x}^{-1/1-q}\left|
1+\chi \right| $. There are two regimes for time evolution at $\mu
_{c}(\sigma )$. When $\tau <\tau _{x}$ the fluctuations are smaller than the
distances between neighboring subsequence positions of the $\sigma =0$ orbit
at $\mu _{c}(0)$, and the iterate positions with $\sigma >0$ fall within
small non overlapping bands each around the $\sigma =0$ position for that $%
\tau $. Time evolution follows a subsequence pattern analogous to that in
the noiseless case. When $\tau \sim \tau _{x}$ the width of the
noise-generated band reached at time $\tau _{x}=2^{N}$ matches the distance
between adjacent positions where $N\sim -\ln \sigma /\ln \kappa $, and this
implies a cutoff in the progress along the position subsequences. At longer
times $\tau >\tau _{x}$ the orbits no longer follow the detailed
period-doubling structure of the attractor. The iterates now trail through
increasingly chaotic trajectories as bands merge with time. This is the
dynamical image - observed along the time evolution for the orbits of a
single state $\mu _{c}(\sigma )$ - of the static bifurcation gap originally
described in terms of the variation of the control parameter $\mu $ \cite
{crutchfield1}, \cite{crutchfield2}, \cite{shraiman1}. The plateau structure
of relaxation and the crossover time $t_{x}$ can be clearly observed in Fig.
1b in Ref. \cite{baldovin4} where $<x_{t}^{2}>-$ $<x_{t}>^{2}$ is shown for
several values of $\sigma $.

\section{Parallels with dynamics near glass formation}

We recall the main dynamical properties displayed by supercooled liquids on
approach to glass formation. One is the growth of a plateau and for that
reason a two-step process of relaxation in the time evolution of two-time
correlations \cite{debenedetti1}. This consists of a primary power-law decay
in time difference $\Delta t$ (so-called $\beta $ relaxation) that leads
into the plateau, the duration $t_{x}$ of which diverges also as a power law
of the difference $T-T_{g}$ as the temperature $T$ decreases to a glass
temperature $T_{g}$. After $t_{x}$ there is a secondary power law decay
(so-called $\alpha $ relaxation) away from the plateau \cite{debenedetti1}.
A second important (nonequilibrium) dynamic property of glasses is the loss
of time translation invariance observed for $T$ below $T_{g}$, a
characteristic known as aging \cite{bouchaud1}. The time fall off of
relaxation functions and correlations display a scaling dependence on the
ratio $t/t_{w}$ where $t_{w}$ is a waiting time. A third notable property is
that the experimentally observed relaxation behavior of supercooled liquids
is effectively described, via reasonable heat capacity assumptions \cite
{debenedetti1}, by the so-called Adam-Gibbs equation, $t_{x}=A\exp
(B/TS_{c}) $, where $t_{x}$ is the relaxation time at $T$, and the
configurational entropy $S_{c}$ is related to the number of minima of the
fluid's potential energy surface \cite{debenedetti1}. We compare the dynamic
properties at the edge of chaos described in the previous section with those
known for the process of vitrification of a liquid as $T\rightarrow T_{g}$.

At noise level $\sigma $ the orbits visit points within the set of $2^{N}$
bands and, as explained in Ref. \cite{robledo3}, this takes place in time in
the same way that period doubling and band merging proceeds in the presence
of a bifurcation gap when the control parameter is run through the interval $%
0\leq \mu \leq 2$. Namely, the trajectories starting at $x_{0}=0$ duplicate
the number of visited bands at times $\tau =2^{n}$, $n=1,...,N$, the
bifurcation gap is reached at $\tau _{x}=$ $2^{N}$, after which the orbits
fall within bands that merge by pairs at times $\tau =2^{N+n}$, $n=1,...,N$.
The sensitivity to initial conditions grows as $\xi _{t}=\exp _{q}(\lambda
_{q}t)$ ($q=1-\ln 2/\ln \alpha <1$) for $t<t_{x}$, but for $t>t_{x}$ the
fluctuations dominate and $\xi _{t}$ grows exponentially as the trajectory
has become chaotic ($q=1$) \cite{robledo3}. This behavior was interpreted 
\cite{robledo3} to be the dynamical system analog of the $\alpha $
relaxation in supercooled fluids. The plateau duration $t_{x}\rightarrow
\infty $ as $\sigma \rightarrow 0$. Additionally, trajectories with initial
conditions $x_{0}$ not belonging to the attractor exhibit an initial
relaxation process towards the plateau as the orbit approaches the
attractor. This is the map analog of the $\beta $ relaxation in supercooled
liquids.

The entropy $S_{c}(\mu _{c}(\sigma ))$ associated to the distribution of
iterate positions (configurations) within the set of $2^{N}$ bands was
determined in Ref. \cite{robledo3}. This entropy has the form $S_{c}(\mu
_{c}(\sigma ))=2^{N}\sigma s$, since each of the $2^{N}$ bands contributes
with an entropy $\sigma s$, where $s=-\int_{-1}^{1}p(\chi )\ln p(\chi )d\chi 
$ and where $p(\chi )$ is the distribution for the noise random variable.
Given that $2^{N}=$ $1+t_{x}$ and $\sigma =(1+t_{x})^{-1/1-r}$, one has $%
S_{c}(\mu _{c},t_{x})/s=(1+t_{x})^{-r/1-r}$or, conversely, 
\begin{equation}
t_{x}=(s/S_{c})^{(1-r)/r}.  \label{adamgibbs1}
\end{equation}
Since $t_{x}\simeq \sigma ^{r-1}$, $r-1\simeq -0.3668$ and $(1-r)/r\simeq
0.5792$ then $t_{x}\rightarrow \infty $ and $S_{c}\rightarrow 0$ as $\sigma
\rightarrow 0$, i.e. the relaxation time diverges as the 'landscape' entropy
vanishes. We interpret this relationship between $t_{x}$ and the entropy $%
S_{c}$ to be the dynamical system analog of the Adam-Gibbs formula for a
supercooled liquid. Notice that Eq.(\ref{adamgibbs1}) is a power law in $%
S_{c}^{-1}$ while for structural glasses it is an exponential in $S_{c}^{-1}$
\cite{debenedetti1}. This difference is significant as it indicates how the
superposition of molecular structure and dynamics upon the bare ergodicity
breakdown phenomenon built in the map modifies the vitrification properties.

The aging scaling property of the trajectories $x_{t}$ at $\mu _{c}(\sigma )$
was examined in Ref. \cite{robledo3}. The case $\sigma =0$ is readily
understood because this property is actually built into the position
subsequences $x_{\tau }=\left| g^{(\tau )}(0)\right| $, $\tau =(2k+1)2^{n}$, 
$k,n=0,1,...$ referred to above. These subsequences can be employed for the
description of trajectories that are at first held at a given attractor
position for a waiting period of time $t_{w}$ and then released to the
normal iterative procedure. For illustrative purposes we select the holding
positions to be any of those for a waiting time $t_{w}=2k+1$, $k=0,1,...$
and notice that for the $x_{in}=0$ orbit these positions are visited at odd
iteration times. The lower-bound positions for these trajectories are given
by those of the subsequences at times $(2k+1)2^{n}$. See Fig. 1. Writing $%
\tau $  as $\tau =$ $t_{w}+t$ we have that $t/t_{w}=2^{n}-1$ and $%
x_{t+t_{w}}=g^{(t_{w})}(0)g^{(t/t_{w})}(0)$ or 
\begin{equation}
x_{t+t_{w}}=g^{(t_{w})}(0)\exp _{q}(-\lambda _{q}t/t_{w}).
\label{trajectory5}
\end{equation}
This fully developed aging property is gradually modified when noise is
turned on. The presence of a bifurcation gap limits its range of validity to
total times $t_{w}+t$ $<t_{x}(\sigma )$ and so progressively disappears as $%
\sigma $ is increased.

\section{Concluding remarks}

The implications of joining the results about intermittency described in
Sections 2 and 3 are apparent. In the critical clusters of infinite size $%
R\rightarrow \infty $ the dynamics of fluctuations obeys the nonextensive
statistics. This is expressed via the time series of the average order
parameter $\Phi _{n}$, i.e. trajectories $\Phi _{n}$ with close initial
values separate in a superexponential fashion according to Eq. (\ref
{sensitivity_00}) with $q=(2\delta +1)/(\delta +1)>1$ and with a $q$%
-Lyapunov coefficient $\lambda _{q}$ determined by the system parameter
values $\delta $, $g_{1}$, $g_{2}$ \cite{robledo2}.

Also, as described in Sections 4 and 5, the dynamics of noise-perturbed
logistic maps at the chaos threshold exhibits the most prominent features of
glassy dynamics in supercooled liquids. The existence of this analogy cannot
be considered accidental since the limit of vanishing noise amplitude $%
\sigma \rightarrow 0$ (the counterpart of the limit $T-T_{g}\rightarrow 0$
in the supercooled liquid) entails loss of ergodicity. Here we have shown
that this nonergodic state corresponds to the limiting state, $\sigma
\rightarrow 0$, $t_{x}\rightarrow \infty $, for a family of small $\sigma $
noisy states with glassy properties, that are noticeably described for $%
t<t_{x}$ via the $q$-exponentials of the nonextensive formalism \cite
{robledo3}.

What is the significance of the connections we have reviewed? Are there
other connections between critical phenomena and transitions to chaos? Are
all critical states - infinite correlation length with vanishing Lyapunov
coefficients - outside BG statistics? Where, and in that case why, does
Tsallis statistics apply? Is ergodicity failure the basic playground for
applicability of generalized statistics? We can mention some noticeable
limitations in the examples discussed. In the case of intermittency we have
focused only on a single laminar episode (or scape from a position very
close to tangency), though this can be of very large duration. In the case
of dynamics at the edge of chaos we have payed attention only to the
dominant features of the multifractal attractor (the most sparse $x=0$ and
most crowded $x=1$ regions) as starting and final orbit positions.

{\bf Acknowledgments.} I am grateful to Constantino Tsallis for many
valuable  discussions. I thank Fulvio Baldovin and Estela Mayoral-Villa for
their exceptional participation in the studies here described. Work
partially supported by DGAPA-UNAM and CONACyT (Mexican Agencies).

\end{document}